# Optimizing spectral distribution character of the LEDs to decrease discoloring of the collections in museum

Chang Ho Kim[1,3,*], Hong Wei Liang[3], Sung Hyok Han[2], Ju Yong Kim[2], Ki Won Ryang[2], Chol Kim[1]

[1]Department of Information Technology, Kim Hyong Jik University, Pyongyang, D. P. R. Korea

[2]Department of Electronics, Kim Chaek University of Technology, Pyongyang, D. P. R. Korea

[3]School of Physics and Optoelectronic Engineering, Dalian University of Technology, Dalian 116024, China

**Abstract**

For white LEDs used for lighting museums, it is possible to reduce their effects on the discoloration of exhibits to a great extent by regulating their spectral distribution so that less lights with 420~470 nm of wavelength which acts on increasing the span of preservation of exhibits, such as pictures, color paper and color cloth. For same illumination of radiation of 5000 lx of white LEDs with different color temperature of about 3000, 3200, 4200 and 6500 K, the density of radiation energy of 420 nm was 34.2, 71.8, 83.1 and 268.3 $\mu W/cm^2$, respectively. The discoloration experiment shows that the effects of discoloration of cold white LEDs was much greater than those of warm white LEDs.

## 1. Introduction

LEDs are being used in the lighting due to its high efficiency and energy saving [5, 9]. In museum, it is important in using LEDs for lighting exhibits that to improve their color rendering index and luminous influence, while increasing the span of preservation of exhibits [1]. Even in rooms without windows, exhibits experience discoloration depending on the density of ultraviolet rays and short wave visible rays emitted from light sources, the air composition and temperature and humidity of rooms [3, 10, 11]. White LEDs with blue lighting emitting diodes coated with yellow phosphor powder are known not to radiate ultraviolet rays so that it is not to have great effects on the discoloration of exhibits generally, but the blue light emitted from LEDs during decades of years of use acts on the discoloration of exhibits to a great extent [6, 7, 8]. However, the amount of blue light is reduced too much, the color rendering index of LED light source and therefore, the estimation value of color rendering index should be kept at over 80 [2, 4].

In this paper, we carried out the discoloration experiments under different LEDs and evaluated the optimizing spectral distribution in order to minimize the discoloration of exhibits by LED lighting in museum.

## 2. Experimental

Various LEDs with different color temperatures and coordinates were manufactured as shown in Table 1 by varying the composition and quantity of fluorescent substances for testing.

---

[*] Corresponding author.
E-mail address: kimchangho@mail.dlut.edu.cn (C.H. Kim).

Table1. Color temperatures and Color coordinates of Sample LEDs

| Color | Color temperature(K) | Color coordinate | |
|---|---|---|---|
| | | x, y | u, v |
| Cold white | 6675 | 0.3085, 0.3356 | 0.1925, 0.3141 |
| White | 5725 | 0.3273, 0.3453 | 0.2017, 0.3193 |
| Pearl red | 4256 | 0.3705, 0.3731 | 0.2200, 0.3323 |
| Warm white | 3224 | 0.4333, 0.4218 | 0.2397, 0.3524 |
| Warm white | 2924 | 0.4612, 0.4433 | 0.2494, 0.3595 |

The size of laboratory for estimating characteristics is 6*6*4 m$^3$ and temperature and humidity are 22±0.5 ℃ and 55±5 RH, respectively. Four square testing boxes were prepared that are painted with black inside with its side is 0.6 m, four types of LEDs-2924 K, 3224 K, 4256 K, and 6675 K were placed on them in the square and adjusted the height of them to illumination intensity of bottom are equal to 5000 lx. We keep the dark room and place their fronts towards the four sides of laboratory to there are no light interference. The spectral distribution curve for of the pre-manufactured cool white LEDs, normal white LEDs, and warm white LEDs were measured; by color electronic panalyzor (HKD-SSP6612) and spectral distribution profile is shown in the figure 1.

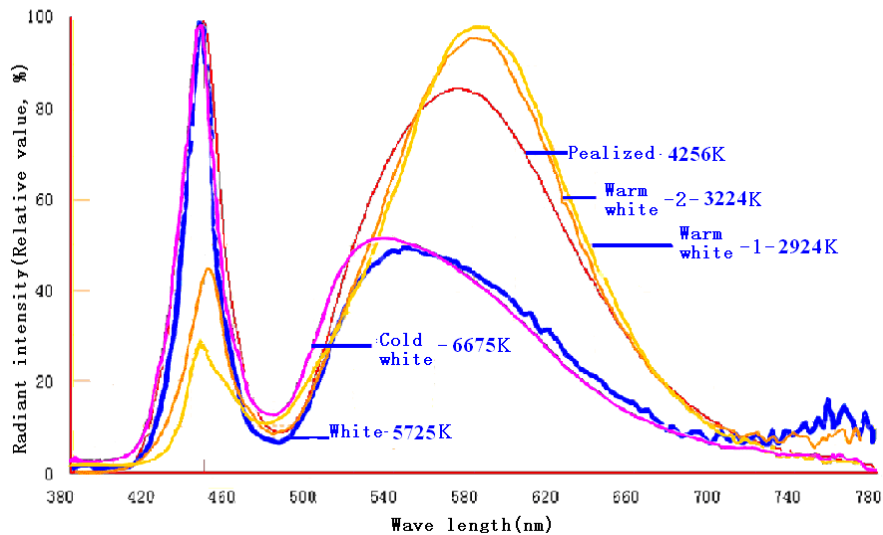

Figure 1．The spectral distribution of the different kinds of LEDs

As shown in figure 1, there is a difference between the intensity of blue light with 420~470nm emitted from the different LEDs. The rate of blue light emission intensity to the total light intensity emitted from cold white LEDs is about 3.1 times that of warm light LEDs (2924 K), 2.7 times that of white LEDs, and 2.03 times that of normal white LEDs. The intensity of blue light at 420 nm was also measured by employing an ultraviolet intensity-meter (MC121817) in the case of 5000 lx.The results show that the density of radiation energy is 34.2 μW/cm$^2$ for LEDs with 2924 K, 71.8 μW/cm$^2$ for those with 3224 K, 83.1 μW/cm$^2$ for those with 4256 K and 268.3 μW/cm$^2$ for those with 6675 K. The density of pearl red LED (4256 K) is 2.4 and 1.2 times that of warm white LEDs (2924 K and 3224 K), respectively, while the density of cold white LED (6675 K) is 7.8 and 3.7 times that of warm white LEDs (2924 K and 3224 K), respectively.

Considering that the blue spectral radiation characteristics is increased according to color temperature，therefore, the blue ratio of cold white LEDs or white LEDs affect the discoloration of exhibits as much.

## 3. Results and discussion

In order to investigate the exhibits decolored characterization influenced by the above LEDs. Experiments were carried out as follows; with a view to evaluating the degree of discoloration of color paper and cloth used for computer-printed plates in museums or exhibition halls, dark measuring rooms and experiment boxes were used, samples of red, orange, yellow, green, blue and violet colors of printing machine coated with anti-discoloration plastic sheet were placed on the floor of these boxes and illumination of 5000 lx was provided on the samples.

Their measurements were 20*30 mm$^2$. The experiments for estimation of discoloration within short time employed accelerated aging test with 5000 lx provided on the samples within the boxes. Illumination intensity is measured by digital illuminometer. Comparison with average museum operation was made on condition that annual energy of 300000 lx·h be incident on the samples assuming that museums will be operated for 5 hours per day, for 300 days per year with their illumination at 200 lx. The measurements for discoloration experiment with 5000 lx for 600 h are equal to $3\times10^6$ lx·h that is the same as the one after 10 years under the normal show condition and at this time, the change of situation of sample colors is clearly visible used for cool white LEDs for lighting. We compare and estimate the situation of discoloration for 10 years with color difference-meter, CR-300 (Made in Japan).

Color difference of color paper and source is calculated by using Hunter's color difference equation [5]

$$\Delta E_{Hunter} = \left[(\Delta L)^2 + (\Delta a)^2 + (\Delta b)^2\right]^{1/2},$$

where $L$ is shade of non-reflecting light, $a$ is expressed by the terms of red and green in reflected light, and $b$ is the terms of yellow and blue in it, $\Delta a$ and $\Delta b$ and $\Delta L$, represents the brightness and color coordinate difference respectively. Color coordinates are calculated as follows.

$$L = 10Y^{1/2},$$

$$a = \frac{17.5\left(\dfrac{X}{f_{XA} + f_{XB}} - Y\right)}{Y^{1/2}},$$

$$\Delta b = \frac{7.0\left(Y - \dfrac{Z}{f_{ZB}}\right)}{Y^{1/2}},$$

where $X$, $Y$ and $Z$ are tristimulus of colored sample and $f_{XA}$, $f_{XB}$ and $f_{ZB}$ are constants by Standard CIE color plate observer and Standard lighting.

This Hunter's color difference equation is color difference computational formula of $L$, $a$ and $b$ systems of opponent-color difference that was reported by Hunter to read the color difference from colorimeter and published in 1948 and it is widely used in the calculating color difference of ceramics, plastic and weaving materials. It can normally meet the demand of industrial production administration. The numbers of table 2-6 are the average ones that is measured 3 times. ΔE, that is the color difference of before and after accelerated testing of discoloration change is calculated by CIE 1976 $L^*a^*b^*$ formula (AATTC1995)

Table 2. Color coordinates of sample color

| Color | Coordinate of the color | |
|---|---|---|
| | Y, x, y | L, a, b |
| Red | 30.64, 0.5168, 0.3713, | 61.95, 41.85, 49.74 |
| Orange | 45.48, 0.4645, 0.4145 | 73.08, 17.55, 57.54 |
| Yellow | 61.43, 0.4410, 0.4445 | 82.51, 1.57, 67.70 |
| Green | 55.32, 0.3296, 0.4520 | 79.07, -38.62, 42.40 |
| Blue | 33.97, 0.2412, 0.2380 | 64.88, 3.80, -31.82 |
| Purple | 28.89, 0.3231, 0.2320 | 60.66, 41.09, -23.13 |

Table 3. Color coordinate difference that is experimented by LED with 6675 K (during 10 years with $3*10^6$ lx·h) of sample color

| Color | Changed coordinate of color | | ΔE |
|---|---|---|---|
| | Y, x, y | L, a, b | |
| Red | 35.61, 0.4751, 0.3695 | 66.00, 33.46, 41.60 | 12.37 |
| Orange | 54.34, 0.4248, 0.4045 | 78.52, 9.40, 47.62 | 13.94 |
| Yellow | 69.44, 0.4103, 0.4285 | 86.89, -3.58, 55.85 | 13.6 |
| Green | 59.31, 0.3109, 0.4057 | 81.27, -33.09, 26.98 | 16.53 |
| Blue | 42.46, 0.2463, 0.2560 | 71.15, -2.39, -27.10 | 9.86 |
| Purple | 36.37, 0.3214, 0.2487 | 66.87, 34.26, -19.26 | 10.01 |

From table 3, 4, 5 and figure 2, the color differences of LED with 6675 K is 13.94, 13.6 and 16.54 and those of LED with 2924 K is 9.56, 6.5 and 14.09 in yellow, green and orange, respectively. The color differences of LED with 2924 K is smaller than 80% of the case LED with 6675 K. Especially, the yellow difference is smaller than half. Therefore, it knows that the lower the temperature of color, the color difference is smaller.

Table 4. Color coordinate difference that is experimented by LED with 3224 K (during 10 years with 3*10⁶lx·h) of sample color

| Color | Changed coordinate of color | | ⊿E |
|---|---|---|---|
| | Y, x, y | L, a, b | |
| Red | 35.08,0.4750,0.3754 | 65.80,31.27,42.80 | 12.14 |
| Orange | 50.71,0.4312,0.4099 | 76.51,9.43,49.53 | 11.73 |
| Yellow | 65.36,0.4180,0.4358 | 84.66,-3.20,59.49 | 9.57 |
| Green | 57.21,0.3135,0.4059 | 80.29,-31.77,27.18 | 16.47 |
| Blue | 39.96,0.2487,0.2554 | 69.44,-0.86,-26.50 | 8.55 |
| Purple | 34.51,0.3249,0.2485 | 65.36,35.27,-18.55 | 8.56 |

Table 5. Color coordinate difference that is experimented by LED with 2924 K (during 10 years with 3*10⁶lx·h) of sample color

| Color | Changed coordinate of color | | ΔE |
|---|---|---|---|
| | Y, x, y | L, a, b | |
| Red | 35.56, 0.4839, 0.3771 | 65.79, 33.65, 45.91 | 9.83 |
| Orange | 52.72, 0.4383, 0.4145 | 77.71, 10.21, 53.52 | 9.56 |
| Yellow | 66.81, 0.4243, 0.4415 | 85.16, -2.84, 63.73 | 6.5 |
| Green | 59.21, 0.3135, 0.4117 | 81.27, -34.10, 29.24 | 14.09 |
| Blue | 41.01, 0.2444, 0.2530 | 70.24, -1.87, -28.12 | 8.64 |
| Purple | 39.39, 0.3197, 0.2475 | 65.96, 34.14, -19.73 | 9.38 |

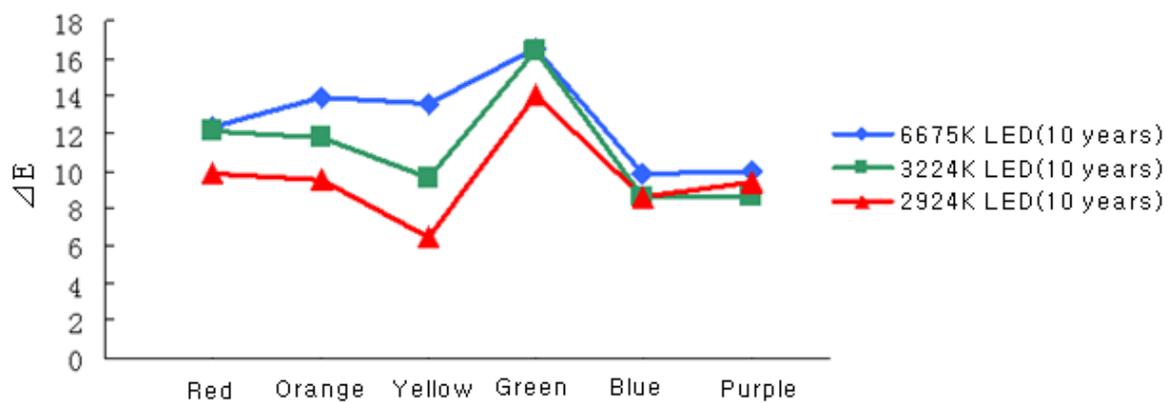

Figure 2. after 10 years, ΔE change of 6675 K LED, 3224 K LED and 2924 K LED

Table 6. Color coordinate difference that is experimented by LED with 6675 K (during 20 years with $6*10^6$lx·h) of sample color

| Color | Changed coordinate of color | | ⊿E |
|---|---|---|---|
| | Y, x, y | L, a, b | |
| Red | 39.68, 0.4523, 0.3675 | 69.24, 28.90, 37.35 | 19.35 |
| Orange | 59.77, 0.4088, 0.3990 | 81.71, 6.16, 43.59 | 19.97 |
| Yellow | 72.92, 0.3997, 0.4213 | 88.40, -4.92, 52.03 | 17.95 |
| Green | 61.19, 0.3086, 0.3883 | 82.47, -28.73, 21.96 | 22.98 |
| Blue | 47.15, 0.2529, 0.2641 | 74.28, -3.07, -24.35 | 13.83 |
| Purple | 41.56, 0.3266, 0.2577 | 70.56, 33.27, -16.28 | 14.36 |

Table 7. Color coordinate difference that is experimented by LED with 2924 K (during 20 years with $3*10^6$lx·h) of sample color

| Color | Changed coordinate of color | | ⊿E |
|---|---|---|---|
| | Y, x, y | L, a, b | |
| Red | 37.56, 0.4740, 0.3737 | 67.76, 32.34, 43.16 | 12.94 |
| Orange | 55.90, 0.4311, 0.4115 | 79.55, 9.16, 51.67 | 12.11 |
| Yellow | 68.72, 0.4213, 0.4384 | 86.31, -2.96, 62.32 | 7.99 |
| Green | 61.00, 0.3111, 0.3938 | 82.38, -29.51, 23.92 | 20.53 |
| Blue | 43.04, 0.2471, 0.2563 | 72.56, -2.13, -27.03 | 10.82 |
| Purple | 38.75, 0.3265, 0.2530 | 68.56, 34.93, -17.54 | 11.47 |

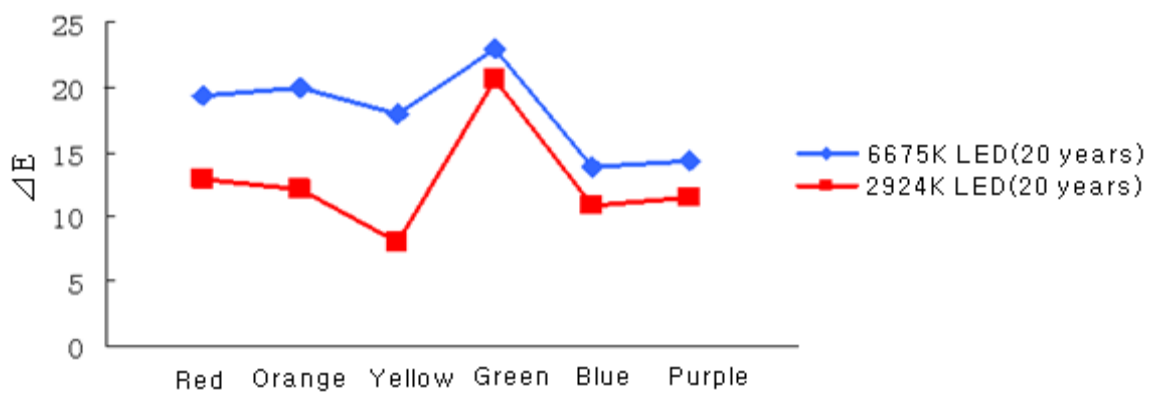

Figure 3. After 20 years, ΔE change of 6675 K LED and 2924 K LED

From table 6, 7 and figure 3, the color difference of samples of red, orange, yellow and green is larger than blue or violet among all the LEDs in the case of warm white LED with 2924 K or cool white LED with 6675 K. Color coordinate changes and color difference of sample colors is shown figure 8-10 when are measured by LEDs with 6675 K, 3224 K and 2924 K for $6.975*10^6$lx·h.

Table 8. Color coordinate difference that is experimented by LED with 6675 K (during 22.5 years with 6.975*10$^6$lx·h) of sample color

| Color | Changed coordinate of color | | ⊿E |
|---|---|---|---|
| | Y, x, y | L, a, b | |
| Red | 41.92,0.4446,0.3669 | 70.81,27.32,36.30 | 22.08 |
| Orange | 60.96,0.4059,0.3970 | 82.35,5.94,42.57 | 21.10 |
| Yellow | 73.53,0.3980,0.4183 | 88.70,-4.54,50.74 | 19.09 |
| Green | 62.98,0.3109,0.3870 | 83.43,-27.60,22.16 | 23.24 |
| Blue | 48.89,0.2561,0.2682 | 75.38,-3.49,-22.81 | 15.64 |
| Purple | 42.50,0.3296,0.2605 | 71.21,33.31,-15.03 | 15.44 |

Table 9. Color coordinate difference that is experimented by LED with 3224 K (during 22.5 years with 6.975*10$^6$lx·h) of sample color

| Color | Changed coordinate of color | | ΔE |
|---|---|---|---|
| | Y, x, y | L, a, b | |
| Red | 40.52,0.4653,0.3726 | 69.83,31.05,41.93 | 15.48 |
| Orange | 58.35,0.4241,0.4072 | 80.93,8.48,49.31 | 14.54 |
| Yellow | 70.44,0.4156,0.4340 | 87.20,-3.52,59.74 | 10.58 |
| Green | 62.77,0.3123,0.3894 | 83.32,-27.79,23.06 | 22.56 |
| Blue | 45.43,0.2523,0.2605 | 73.17,-1.60,-25.39 | 13.27 |
| Purple | 39.68,0.3328,0.2540 | 69.23,37.27,-16.50 | 13.20 |

Table 10. Color coordinate difference that is experimented by LED with 2924 K (during 22.5 years with 6.975*10$^6$lx·h) of sample color

| Color | Changed coordinate of color | | ΔE |
|---|---|---|---|
| | Y, x, y | L, a, b | |
| Red | 37.29,0.4712,0.3762 | 67.49,30.58,43.18 | 14.30 |
| Orange | 57.23,0.4286,0.4138 | 80.30,7.61,52.24 | 13.27 |
| Yellow | 69.34,0.4189,0.4394 | 86.67,-4.18,62.21 | 8.96 |
| Green | 62.24,0.3118,0.3947 | 83.04,-29.71,24.44 | 20.56 |
| Blue | 45.47,0.2498,0.2615 | 73.19,-3.38,-25.35 | 12.70 |
| Purple | 39.41,0.3281,0.2561 | 69.04,34.17,-16.36 | 12.61 |

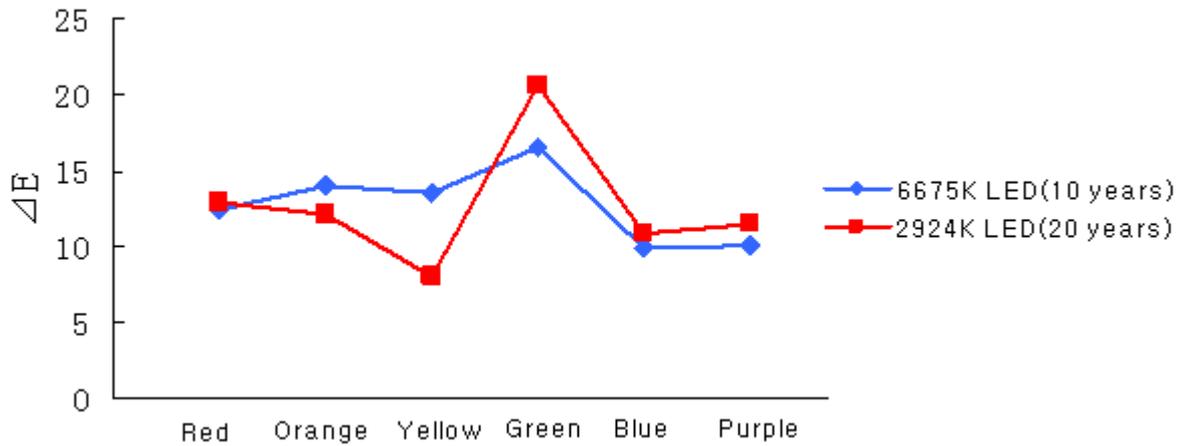

Figure 4. ΔE change of 6675 K LED (after 10 years) and 2924 K LED (after 20 years)

Comparing table 3 and 7 (figure 4), after 20 years, the color differences of LED with 2924 K is 8.96 and 13.27 in yellow and orange, respectively, but after 10 years, that is 13.94 and 13.6. Therefore, we can see that discoloration of LED with 2924 K for 20 years is much smaller than the LED with 6675 K for 10 years. As shown in figure 3, for the LED with 2924 K after 20 years, color differences of orange and yellow are smaller than the LEDs with 6675 K after 10 years and red, ones of red, blue and violet is a bit larger than them.

However, such changes have no effects on the exhibitions. Estimating the testing samples of real red, orange, yellow, green, blue and violet that is employed by accelerated aging test in the view of artist, six color samples give cool feeling because warm colors are changed. So for the exhibitions painted by these six colors, they can lose their values to change their colors.

Using the data of table 3-10, discoloration profile (color difference value) is shown as figure 4-6 depending on time (for years)

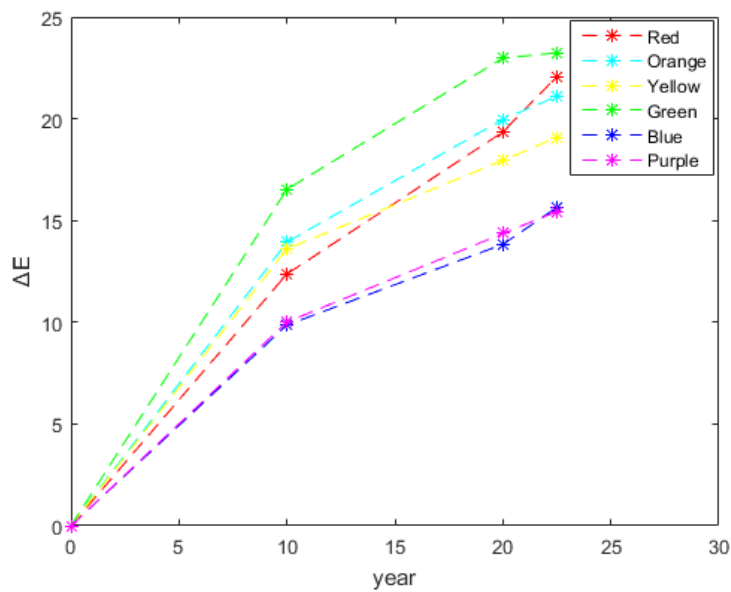

Figure 5. Discoloration Profile that is experimented by LEDs with 6675 K depending on time

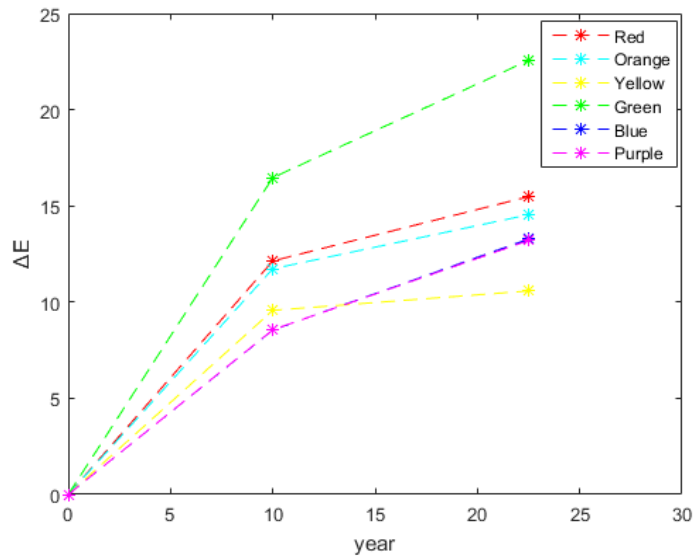

Figure 6. Discoloration Profile that is experimented by LEDs with 3224 K depending on time

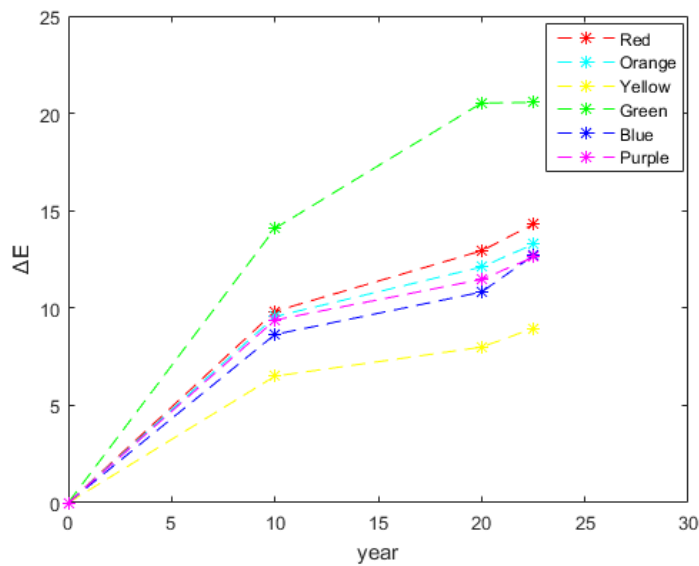

Figure 7. Discoloration Profile that is experimented by LEDs with 2924 K depending on time

As shown figure 5, 6, 7 for all the cases, gradients of curves express the discoloration speed so it is fast for the first ten years and speed is getting slowly as time goes by. And in the case of cool while LEDs (6675 K) is larger than warm white LEDs.

### 4. Conclusion

The factors that is high effect on the discoloration of exhibits are illumination intensity and spectral distribution characteristics of LEDs for lighting, in detail they are illumination intensity and blue peak radiative quantity. In this paper, we estimated that warm white LEDs whose quantity of light of blue peak wave length is decreased 3 times have less effects than white or cool white LEDs on the discoloration of exhibitions such as includes computer printing pictures, color paper and color cloth. Discoloration of exhibits is happened by blue radiation with big energy for 420-470 nm and according to the color temperature the rate of blue radiation in all the light beam is different so discoloration is large as this rate. So for illuminating museums, it is good to use warm white LEDs with 3000K that blue radiation is relatively small.